\documentstyle[12pt]{article}

\catcode`\@=11
\long\def\@makefntext#1{
\protect\noindent \hbox to 3.2pt {\hskip-.9pt
$^{{\ninerm\@thefnmark}}$\hfil}#1\hfill}		

\def\@makefnmark{\hbox to 0pt{$^{\@thefnmark}$\hss}}  

\def\ps@myheadings{\let\@mkboth\@gobbletwo
\def\@oddhead{\hbox{}
\rightmark\hfil\ninerm\thepage}
\def\@oddfoot{}\def\@evenhead{\ninerm\thepage\hfil
\leftmark\hbox{}}\def\@evenfoot{}
\def\sectionmark##1{}\def\subsectionmark##1{}}

\renewcommand{\thefootnote}{\fnsymbol{footnote}}

\newcounter{sectionc}\newcounter{subsectionc}\newcounter{subsubsectionc}
\renewcommand{\section}[1] {\vspace*{0.6cm}\addtocounter{sectionc}{1}
\setcounter{subsectionc}{0}\setcounter{subsubsectionc}{0}\noindent
	{\normalsize\bf\thesectionc. #1}\par\vspace*{0.4cm}}
\renewcommand{\subsection}[1] {\vspace*{0.6cm}\addtocounter{subsectionc}{1}
	\setcounter{subsubsectionc}{0}\noindent
	{\normalsize\it\thesectionc.\thesubsectionc. #1}\par\vspace*{0.4cm}}
\renewcommand{\subsubsection}[1]
{\vspace*{0.6cm}\addtocounter{subsubsectionc}{1}
	\noindent {\normalsize\rm\thesectionc.\thesubsectionc.\thesubsubsectionc.
	#1}\par\vspace*{0.4cm}}

\newcounter{appendixc}
\newcounter{subappendixc}[appendixc]
\newcounter{subsubappendixc}[subappendixc]

\renewcommand{\appendix}[1] {\vspace*{0.6cm}
        \refstepcounter{appendixc}
        \setcounter{figure}{0}
        \setcounter{table}{0}
        \setcounter{equation}{0}
        \renewcommand{\thefigure}{\Alph{appendixc}.\arabic{figure}}
        \renewcommand{\thetable}{\Alph{appendixc}.\arabic{table}}
        \renewcommand{\theappendixc}{\Alph{appendixc}}
        \renewcommand{\theequation}{\Alph{appendixc}.\arabic{equation}}
        \noindent{\bf Appendix \theappendixc #1}\par\vspace*{0.4cm}}

\def\abstracts#1{{

\centering{\begin{minipage}{12.2truecm}\footnotesize\baselineskip=12pt\noindent
	\centerline{\footnotesize ABSTRACT}\vspace*{0.3cm}
	\parindent=0pt #1
	\end{minipage}}\par}}

\renewenvironment{thebibliography}[1]
	{\begin{list}{\arabic{enumi}.}
	{\usecounter{enumi}\setlength{\parsep}{0pt}
\setlength{\leftmargin 1.25cm}{\rightmargin 0pt}
	 \setlength{\itemsep}{0pt} \settowidth
	{\labelwidth}{#1.}\sloppy}}{\end{list}}

\newcounter{itemlistc}
\newcounter{romanlistc}
\newcounter{alphlistc}
\newcounter{arabiclistc}

\newcommand{\fcaption}[1]{
        \refstepcounter{figure}
        \setbox\@tempboxa = \hbox{\footnotesize Fig.~\thefigure. #1}
        \ifdim \wd\@tempboxa > 6in
           {\begin{center}
        \parbox{6in}{\footnotesize\baselineskip=12pt Fig.~\thefigure. #1}
            \end{center}}
        \else
             {\begin{center}
             {\footnotesize Fig.~\thefigure. #1}
              \end{center}}
        \fi}

\newcommand{\tcaption}[1]{
        \refstepcounter{table}
        \setbox\@tempboxa = \hbox{\footnotesize Table~\thetable. #1}
        \ifdim \wd\@tempboxa > 6in
           {\begin{center}
        \parbox{6in}{\footnotesize\baselineskip=12pt Table~\thetable. #1}
            \end{center}}
        \else
             {\begin{center}
             {\footnotesize Table~\thetable. #1}
              \end{center}}
        \fi}

\def\@citex[#1]#2{\if@filesw\immediate\write\@auxout
	{\string\citation{#2}}\fi
\def\@citea{}\@cite{\@for\@citeb:=#2\do
	{\@citea\def\@citea{,}\@ifundefined
	{b@\@citeb}{{\bf ?}\@warning
	{Citation `\@citeb' on page \thepage \space undefined}}
	{\csname b@\@citeb\endcsname}}}{#1}}

\newif\if@cghi
\def\cite{\@cghitrue\@ifnextchar [{\@tempswatrue
	\@citex}{\@tempswafalse\@citex[]}}
\def\citelow{\@cghifalse\@ifnextchar [{\@tempswatrue
	\@citex}{\@tempswafalse\@citex[]}}
\def\@cite#1#2{{$\null^{#1}$\if@tempswa\typeout
	{IJCGA warning: optional citation argument
	ignored: `#2'} \fi}}

 1
 1
 1

\font\ninerm=cmr9

\begin{document}

\centerline{\normalsize\bf CONFINING BETHE--SALPETER EQUATION IN QCD}

\centerline{\footnotesize N. BRAMBILLA\footnote{Presenting author}
 and G. M. PROSPERI}
\baselineskip=13pt
\centerline{\footnotesize\it Dipartimento di Fisica dell'Universit\`a, Milano}
\baselineskip=12pt
\centerline{\footnotesize\it INFN, Sezione di Milano, Via Celoria 16, 20133
 Milano}
\centerline{\footnotesize E-mail: n.brambilla@mi.infn.it}

\vspace*{0.9cm}
\abstracts{We derive a  confining $ q \bar{q}$ Bethe--Salpeter
 equation starting from the same assumptions on the Wilson loop
 integral already adopted in the derivation of a semirelativistic
 heavy quark potential.  We show that, by standard approximations,
 an effective meson squared mass operator can be obtained  from
our BS kernel and that, from this, by ${1\over m^2}$ expansion, the
corresponding Wilson loop potential is recovered, spin--dependent and
velocity--dependent terms included. We also show, that, on the contrary,
neglecting  spin--dependent terms, relativistic flux tube model is reproduced.}

\normalsize\baselineskip=15pt
\setcounter{footnote}{0}
\renewcommand{\thefootnote}{\alph{footnote}}
\vskip 0.5truecm
In the paper presented by G.M. Prosperi
 ~\cite{prosp} it was shown how the properties
 of the Wilson loop integral
( we assume Wilson area law and   the straight line
approximation; see eqs. (2)--(4))
can be used to obtain a confining Bethe--Salpeter equation
from first principles. This result was accomplished neglecting
the spin of the quarks.
In this paper we show that  it can be extended to
the case  of regular QCD with quarks with spin
 by defining an appropriate operator for the spin dependent part
 and using a second order formalism.\par
Even in this case the basic object
 is the
 quark--antiquark Green function
\begin{eqnarray}
G_4(x_1,x_2,y_1,y_2) &=&
\frac{1}{3}\langle0|{\rm T}\psi_2^c(x_2)U(x_2,x_1)\psi_1(x_1)
\overline{\psi}_1(y_1)U(y_1,y_2)  \overline{\psi}_2^c(y_2)
|0\rangle =
\nonumber\\
&=& \frac{1}{3} {\rm Tr} \langle U(x_2,x_1)
 S_1(x_1,y_1;A) U(y_1,y_2)
\tilde{S}_2(y_2,x_2;-\tilde{A})  \rangle
\label{eq:gauginv}
\end{eqnarray}
where $c$ denotes the charge-conjugate fields,
 $U$
the path-ordered gauge string
$U(b,a)= {\rm P}_{ba}  \exp  \left(ig\int_a^b dx^{\mu} \, A_{\mu}(x) \right)$,
 $S_1$ and $S_2$ the quark propagators in the
external gauge field $A^{\mu}$, the tilde denotes transposition in the colour
 indeces.\par
Then, putting ($D_\nu =\partial_\nu - i g A_\mu $)
$S(x,y;A) = ( i \gamma^\nu D_{\nu} + m) \Delta^\sigma(x,y,;A)$,
we have ($\sigma^{\mu \nu} = {i\over 2} [\gamma^\mu,\gamma^\nu ]$)
\begin{equation}
(D_\mu D^\mu +m^2 -{1\over 2}g \sigma^{\mu \nu} F_{\mu \nu}
)\Delta^\sigma(x,y;A) = - \delta^4 (x-y),
\label{eq:itnov}
\end{equation}
 and taking into account gauge invariance, we can write
\begin{equation}
G_4^{\rm gi}(x_1,x_2; y_1,y_2) =(i \gamma_1^\mu \partial_{x_1 \mu}
 + m_1) ( i \gamma_2^\nu \partial_{x_2 \nu} +m_2) H_4(x_1,x_2;y_1,y_2)
\label{eq:eqg}
\end{equation}
with
\begin{equation}
H_4(x_1,x_2;y_1,y_2) = -{1\over 3} {\rm Tr}
\langle U(x_2,x_1) \Delta^\sigma_1(x_1,y_1;A)
 U(y_1,y_2) \tilde{\Delta}^\sigma_2(x_2,y_2;-\tilde{A})\rangle
\label{eq:eqh}
\end{equation}
Now, we use  the explicit resolution of (\ref{eq:itnov}) in terms
 of a path integral
(Feynman-Schwinger representation; see (14) of Ref.~\cite{prosp})
\begin{eqnarray}
& & \Delta^\sigma (x,y;A)  =
- {i\over 2}\int_0^\infty ds \exp {i s\over 2}
 ( - D_\mu D^\mu - m^2 + {1\over 2} g \sigma^{\mu \nu} F_{\mu \nu})
 \\
& & = -{i\over 2} \int_0^\infty d s
 \int_y^x {\cal D}z\,
  {\rm P}_{xy} {\rm T}_{xy} {\rm exp}\, i \int_0^{s}
 d\tau \{-{1\over 2} (m^2 +\dot{z}^2) + g A_\rho (z) \dot{z}^\rho
 + {g\over 4} \sigma^{\mu \nu} F_{\mu \nu}(z) \}\nonumber
\label{eq:part}
\end{eqnarray}
where the path integral is understood to be extended over all
 paths
 $z^{\mu}= z^{\mu}(\tau)$ connecting $y$ with $x$ and
expressed in
terms of a   parameter $ \tau$
with $ 0\le \tau
\le s$,
$\dot{z}$  stands
for ${dz(\tau)
\over d \tau}$, the ``functional measure'' is assumed to be
defined as
$ {\cal D} z   =  ({1\over 2 \pi i \varepsilon })^{ 2 N}
  d^4 z_1 \dots  d^4 z_{N-1}$,
   ${ \rm P}_{xy}$ and ${\rm T}_{xy}$  prescribe the ordering
along the path from right to left
respectively of the colour matrices  and of the spin matrices.
\par
On the other side, it is well known
that, as a consequence of
 a variation in the path $z^\mu(\tau)
\to z^\mu(\tau)+ \delta z^\mu(\tau)$
respecting the extreme points, one has
\begin{eqnarray}
\delta & &\big \{ {\rm P}_{xy} \exp ig \int_0^s d\tau \dot{z}^\mu(\tau) A_{\mu}
(z)\big \}=\nonumber \\
& & = ig \int_0^s \delta S^{\mu\nu}(z(\tau)) {\rm P}_{xy}
\big \{- F_{\mu \nu}(z(\tau)) \exp ig \int_0^s d\tau^\prime
\dot{z}^\mu(\tau^\prime) A_{\mu}(z(\tau^\prime)) \big \}
\label{eq:varwils}
\end{eqnarray}
with $\delta S^{\mu \nu}(z)= {1\over 2} (d z^\mu \delta z^\nu- dz^\nu
\delta z^\mu)$.
Then
\begin{eqnarray}
{\rm T}_{xy}& &  \exp(-{1\over 4} \int_0^s d\tau \sigma^{\mu\nu} {\delta
\over \delta S^{\mu \nu}(z) } )
\Big ( {\rm P}_{xy} \exp ig \int_0^s d \tau^\prime \dot{z}^\mu(\tau^\prime)
A_{\mu}(z(\tau^\prime)) \Big )\nonumber \\
& & = {\rm T}_{xy} {\rm P}_{xy} \exp ig \int_0^s d\tau [
\dot{z}^\mu(\tau)A_{\mu}(z(\tau)) + {1\over 4} \sigma^{\mu \nu}
F_{\mu \nu}(z(\tau )) ]
\label{eq:deffmu}
\end{eqnarray}
 and
Eq.(\ref{eq:part}) can  be rewritten as
\begin{equation}
\Delta^\sigma(x,y; A)= -{i\over 2} \int_0^s d\tau \int_y^x {\cal
D}z  {\rm P}_{x y}
{\rm T}_{xy} {\cal S}_0^s \exp \, i \int_0^s d\tau [-{1\over 2} (m^2
+\dot{z}^2)
+ ig\dot{\bar{z}}^\mu A_{\mu}(\bar{z})]
\label{eq:partbis}
\end{equation}
with
\begin{equation}
{\cal S}_0^s =  \exp \Big [ -{1\over 4} \int_0^s d \tau
 \sigma^{\mu \nu} {\delta \over \delta S^{\mu \nu}(\bar{z})}
\Big ]
\label{eq:defop}
\end{equation}
In (9)
 it is understood that $ \bar{z}^\mu(\tau)$ has to be put
equal to  $z^\mu(\tau)$ after the action of ${\cal S}_0^{s} $.
 Alternatively, it is convenient
to write $\bar{z}= z+\zeta$, assume that
${\cal S}_0^s$ acts on $\zeta (\tau)$ with
$ \delta S^{\mu \nu} (z) = {1\over 2} ( d z^\mu \delta \zeta^\nu
 - d z^\nu \delta \zeta^\mu )$, and set eventually $\zeta =0$.
\par  Replacing (\ref{eq:partbis}) in (\ref{eq:eqh}) we obtain
\begin{eqnarray}
H_4(x_1,x_2;y_1,y_2) & & = ({1 \over 2})^2 \int_0^{\infty} d s_1
\int_0^{\infty} d s_2
 \int_{y_1}^{x_1} {\cal D} z_1\int_{y_2}^{x_2} {\cal D} z_2
 {\rm T}_{x_1 y_1} {\rm T}_{x_2 y_2}
 {\cal S}_0^{s_1} {\cal S}_0^{s_2} \nonumber \\
& & \exp{ ({-i \over 2}) \big \{
 \int_0^{s_1} d\tau_1 (m_1^2 +\dot{z}_1^2) + \int_0^{s_2}
d\tau_2 (m_2^2 +\dot{z}_2^2)\big \} }\nonumber \\
& & {1\over 3} \langle {\rm Tr} {\rm P}_\Gamma
 \exp (ig) \big \{ \oint_{\Gamma} d\bar{z}^\mu A_{\mu}
(\bar{z})   \}
\rangle
\label{eq:hquatr}
\end{eqnarray}
where now $ \bar{z}= \bar{z}_j= z_j +\zeta_j $ on $\Gamma_1 $ and
$\Gamma_2$ and $ \bar{z}=z $ on the end lines $x_1 x_2$ and $y_2 y_1$;
  the final limit $\zeta_j \to 0$ being again  understood.\par
Eq. (10) corresponds to Eq. (15) of Ref.~\cite{prosp}.
Then, by using assumption (2)--(4) of Ref.~\cite{prosp} for the Wilson loop
integral, the recurrence identity (19)~\cite{prosp} and proceeding  in
a similar way (apart from some technical complications)
we can show~\cite{bs}
that the "second order" Green function $H_4(x_1,\,x_2;\,y_1,
\,y_2)$ satisfies a Bethe-Salpeter type nonhomogeneous equation.
{}From this we obtain
the momentum space homogeneous equation
 \begin{equation}
     \Phi_P (k^\prime) = -i \int {d^4 k \over (2 \pi)^4} \hat H_2(\eta_1 P
+ k^\prime) \hat H_2(\eta_2 P - k^\prime) \hat I(k^\prime , k;P) \Phi_P
(k)
\label{eq:fi}
\end{equation}
which is more appropriate for the bound state problem. In this equation
$H_2$ stands for a colour independent one particle dressed propagator,$
 \eta_j = {m_j \over m_1 + m_2} $, $ P $ denotes
 the total momentum $p_1 + p_2$,
$ k$  the relative momentum $\eta_2 p_1 - \eta_1 p_2 $
($q_j = \eta_j P + { k+k^\prime \over 2}$ and in the CM frame $ {\bf
q} = { {\bf k}^\prime + {\bf k} \over 2}$ ), $\Phi_P(k) $ is
 the ordinary Bethe--Salpeter wave function  and
 \begin{eqnarray}
  & & \hat{I}_{\rm pert} =  16 \pi {4 \over 3} \alpha_s
   \{ D_{\rho \sigma}(Q) q_1^\rho
        q_2^\sigma - {i \over 4} \sigma_1^{\mu \nu} (\delta_\mu^\rho Q_\nu-
     \delta_\nu^\rho Q_\mu) q_2^\sigma D_{\rho \sigma } (Q) \nonumber\\
& & + {i \over 4} \sigma_2^{\mu \nu}
        (\delta_\mu^\sigma Q_\nu - \delta_\nu^\sigma Q_\mu) q_1^\rho
  D_{\rho \sigma }(Q)
+{1 \over 16} \sigma_1^{\mu_1 \nu_1} \sigma_2^{\mu_2 \nu_2}
        (\delta_{\mu_1}^\rho Q_{\nu_1} - \delta_{\nu_1}^\rho Q_{\mu_1})
     \times\nonumber \\
& & \quad \quad   (\delta_{\mu_2}^\sigma Q_{\nu_2} - \delta_{\nu_2}^\sigma
Q_{\mu_2})
        D_{\rho \sigma}(Q) \}  +\dots
\label{eq:ipertq}
\end{eqnarray}
$     \hat{I}_{\rm conf} = \int d^3 {\bf r} \, e^{i {\bf Q} \cdot {\bf r}}\,
     J({\bf r}, \, q_1, \, q_2)$
with
 \begin{eqnarray}
   J({\bf r}, \, q_1, \, q_2) & & = {2 \sigma r \over q_{10} + q_{20} }
    \Big [  q_{20}^2 \sqrt{q_{10}^2  -{\bf q}_{\rm T}^2} +
      q_{10}^2 \sqrt  {q_{20}^2 - {\bf q}_{\rm T}^2} + \nonumber \\
     & &+ {q_{10}^2 q_{20}^2 \over \vert {\bf q}_{\rm T} \vert }
      (\arcsin {\vert {\bf q}_{\rm T}\vert \over  q_{10} } +
      \arcsin {\vert {\bf q}_{\rm T}\vert \over  q_{20}  })
\Big ]
+\nonumber \\
& & +2 \sigma  {\sigma^{k \nu}_1 q_{20} q_{1 \nu} r^k \over r \sqrt{q_{10}^2
-
 {\bf q}_{ {\rm T}}^2}} + 2 \sigma
 { \sigma^{k \nu}_2 q_{10} q_{2 \nu} r^k
\over  r \sqrt{q_{20}^2 - {\bf q}^2_{ {\rm T}}} }
+\dots
\label{eq:iconfj}
\end{eqnarray}
In (\ref{eq:ipertq})--(\ref{eq:iconfj}) we have set $
    q_1={p_1^\prime + p_1 \over 2}$, $ q_2={p_2^\prime
      + p_2 \over 2} $,
     $ Q = p_1^\prime - p_1   = p_2 - p_2^\prime $,
 $D_{\rho \sigma}(Q)$ denotes the gluon free
propagator and
the center of mass system
 ($ {\bf q}_1 = -{\bf q}_2 = {\bf q} \, , \ q_{\rm T}^h =
(\delta^{hk} - \hat r ^h \hat r ^k) q^k $) is understood.
   Notice that
 Eq. (\ref{eq:ipertq}) corresponds to the
usual ladder approximation in this second order  formalism
(differing from (17)~\cite{prosp} only for the spin dependent terms.\par
{}From (\ref{eq:fi}) by replacing $\hat{H}_2 (p)$  with the free propagator
 $ {-i\over p^2 -m^2}  $ and performing an appropriate instantaneous
 approximation on $\hat{I}$ [consisting in setting
$ Q_0=0$, $q_{j0}= { w_j^\prime + w_j\over 2} $ or $
p_{j0}=p_{j0}^\prime ={ w_j^\prime + w_j\over 2} $ or
 $ k_0=k_0^\prime =\eta_2 { w_1^\prime + w_1 \over 2} -\eta_1
{w_2^\prime + w_2\over 2}$ and $ P_0 ={1\over 2} ( w_1^\prime
+ w_1 + w_2^\prime + w_2 )$, with $w_j=\sqrt{m_j^2 + {\bf k}^2 }$,
 $ w_j^\prime=\sqrt{m_j^2 + {\bf k}^{ \prime 2}}$ ]
  one can obtain ~\cite{bs} an
effective  mass operator for the mesons (in the CM frame ${\bf
P}=0, P=(m_B,0) $)
$        M = M_0 + V$
with
\begin{equation}
\langle {\bf k}^\prime \vert V \vert {\bf k} \rangle =
{1 \over ( 2 \pi)^3 }
 {1 \over 4 \sqrt{ w_1 w_2 w_1^\prime w_2^\prime } } \hat{I}_{\rm inst}
({\bf k}^\prime, {\bf k})+\dots
\label{eq:linrel}
\end{equation}
where the dots stand for higher order terms in $\alpha _{\rm s}$ and $\sigma
a^2$  and kinematical factors equal to 1 on the energy shell have been
 neglected.
 Now, if we neglect in $V$ the spin dependent terms and the
coulombian one, we
 reobtain the hamiltonian of the relativistic flux tube
model ~\cite{flux}
 with an appropriate ordering prescription ~\cite{BCP}.
 On the other side by performing a ${1\over m^2}$ expansion
 we find the $ q \bar{q}$ potential that, written in the representation
 space, reads
\begin{eqnarray}
V & & = {4\over 3} {\alpha_s \over r} + \sigma r
+ {4 \over 3} {\alpha_s \over m_1 m_2} \big \{ {1 \over 2 r}
 (\delta^{hk} + {r^h r^k \over r^2 }) q^h q^k \big \}_{\rm
W}\nonumber \\
& & - {4\over 3} i \alpha_s ( {1\over 2 m_1 } {\alpha_1 \cdot {\bf r}
 \over r^3 } - {1\over 2 m_2 } { \alpha_2 \cdot {\bf r} \over r^3})
+ {4\over 3} {\alpha_s \over 2 m_1 m_2} ({\bf \sigma}_1 + {\bf
\sigma}_2 ) \cdot ( {\bf r} \times {\bf q} )\nonumber \\
& & + {1\over 3} {\alpha_s \over m_1 m_2 } [ { 3 ( \sigma_1 \cdot {\bf
r} ) (\sigma_2 \cdot {\bf r}) \over r^5 } - {\sigma_1 \cdot \sigma_2
\over r^3  } ] + {4\over 3} {\alpha_s \over m_1 m_2 }
 { 2 \pi \over 3} (\sigma_1 \cdot \sigma_2) \delta^3({\bf r})\nonumber
\\
& & - {\sigma \over 6} ( {1\over m_1^2} + {1\over m_2^2} -{1\over m_1
m_2} ) \{ {\bf q}^2_{\rm T} r \}_{\rm W}\nonumber \\
& & - {\sigma \over 2} ( {\sigma_1 \over m_1^2 } +{\sigma_2 \over
m_2^2} ) \cdot ( {{\bf r} \over r} \times {\bf q} ) - {\sigma i
\over 2} [ {1\over m_1}  {\alpha_1 \cdot {\bf r} \over r} - {1\over
m_2} { \alpha_2 \cdot {\bf r} \over r} ]
\end{eqnarray}
where now ${\bf q}$ stands for the momentum operator and
and the symbol $\{ \,\,\}_{\rm
W}$ stands for the Weyl ordering
prescription for momentum  and position variables.
Now, by  performing a Foldy--Wouthuysen  tranformation with  generator
$
S= {i \over 2 m_1} \alpha_1 \cdot {\bf q} - {i\over 2 m_2}
 \alpha_2 \cdot {\bf q} $
we end up with the ${1\over m^2}$ potential
 which coincides with the Wilson loop
potential ~\cite{BCP}.\par

\section {Conclusions}
The kernel was constructed as an expansion in $\alpha_{\rm s}$ and $\sigma
a^2$ and at the lowest order is given by equations (9)-(10).\par
 As the analysis in terms of potentials show, the inclusion
of terms in $\alpha_{\rm s}^2$ is essential for an understanding of the fine
and the hyperfine structure.
For what concerns the importance of $\sigma^2$
contributions some preliminary extimates~\cite{bs}
 performed in the relativistic
 flux tube context seem to indicate that this first  correction
 should be of little  significance for the spectrum in almost all cases.
 \par
Finally  let us come to the problem
 of the type of confinement,
 which has been largely discussed in
the literature.  By this terminology it is usually meant  the
tentative  assumption of a BS (first order) confining kernel of the
instantaneous form $
\hat{I}_{\rm conf}= -(2 \pi )^3 \Gamma {\sigma \over \pi^2 } {1\over
 {\bf Q}^4}$.
As well known, the above form of $I$  with $\Gamma=1$ was motivated by the
fact that it reproduces the static potential  $\sigma r$ and the spin
dependent potential as obtained in the Wilson loop  context. This
choice, however, gets both into phenomenological and theoretical
 difficulties:
1) it gives a first order velocity dependent relativistic
correction to the potential which differs from the Wilson loop
 one ~\cite{BCP} and does not seem to agree with the heavy
meson  data ~\cite{guptanoi}; 2)
  it does not reproduce straight line Regge trajectories
 ~\cite{durand,flux}.
Complementary objections can be moved to the form with
 $\Gamma = \gamma_1^0 \gamma_2^0 $.\par
On the contrary, even if we have not yet attempted calculations
directly with the kernel established in this paper, very encouraging
 results have been obtained in the context of the relativistic
 flux tube model ~\cite{flux}, of the dual QCD ~\cite{baker} and
 of the effective
 relativistic  hamiltonian ~\cite{sim}, formalisms that are all
 strictly related to our one. Therefore the complicated momentum
dependence appearing  in (10) seems essential to understand
both the light and the heavy meson phenomenology.

\end{document}